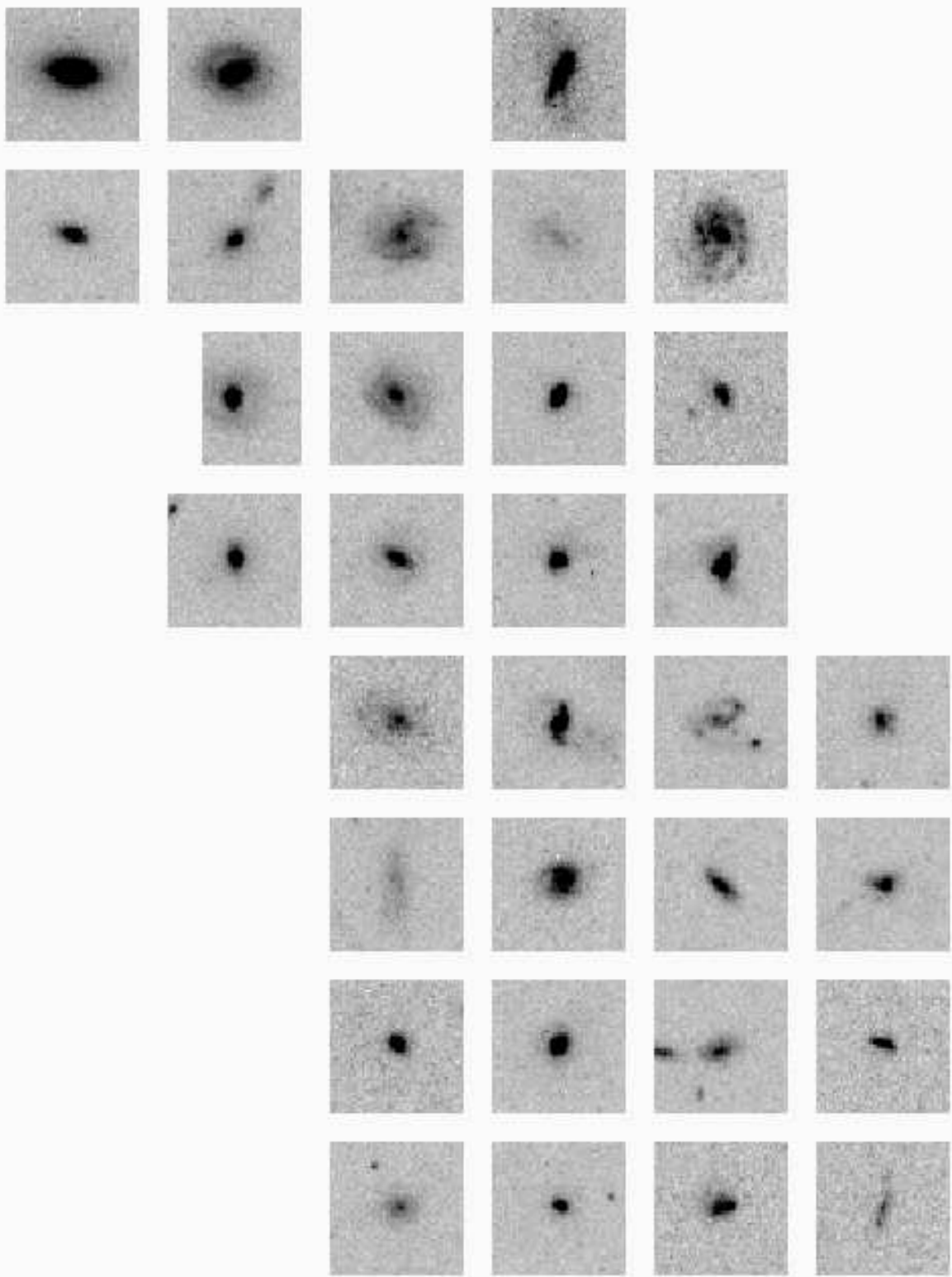

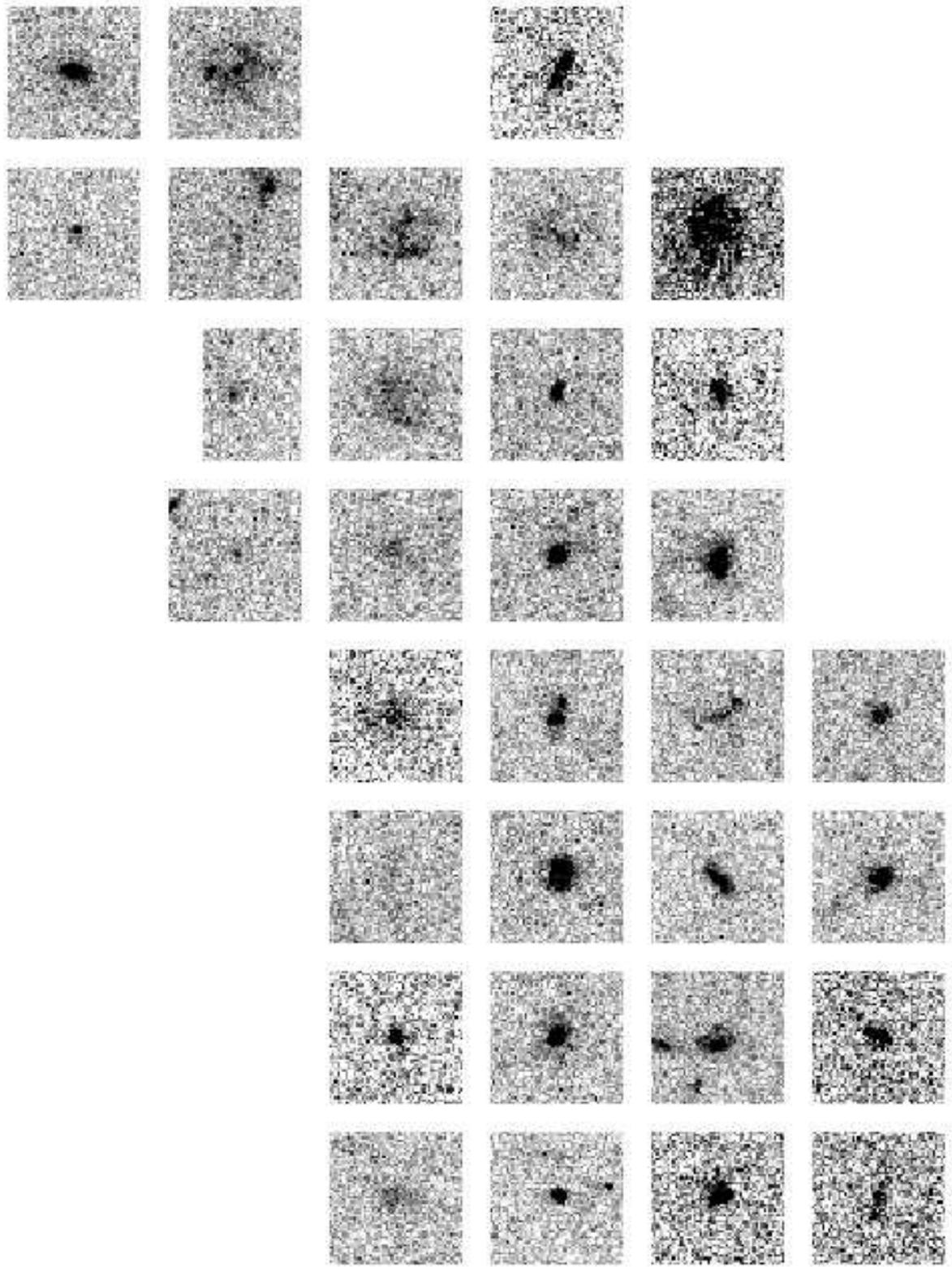

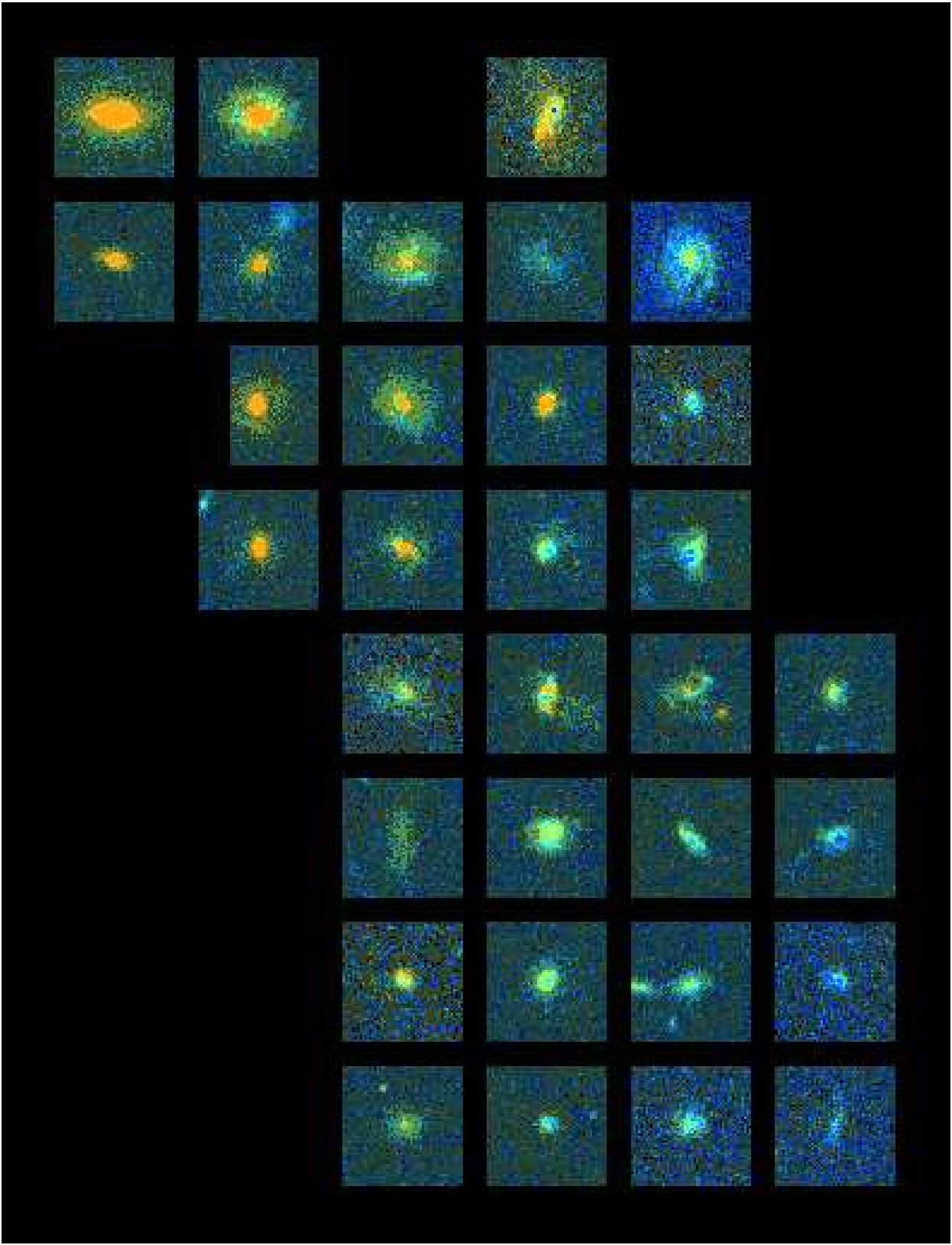

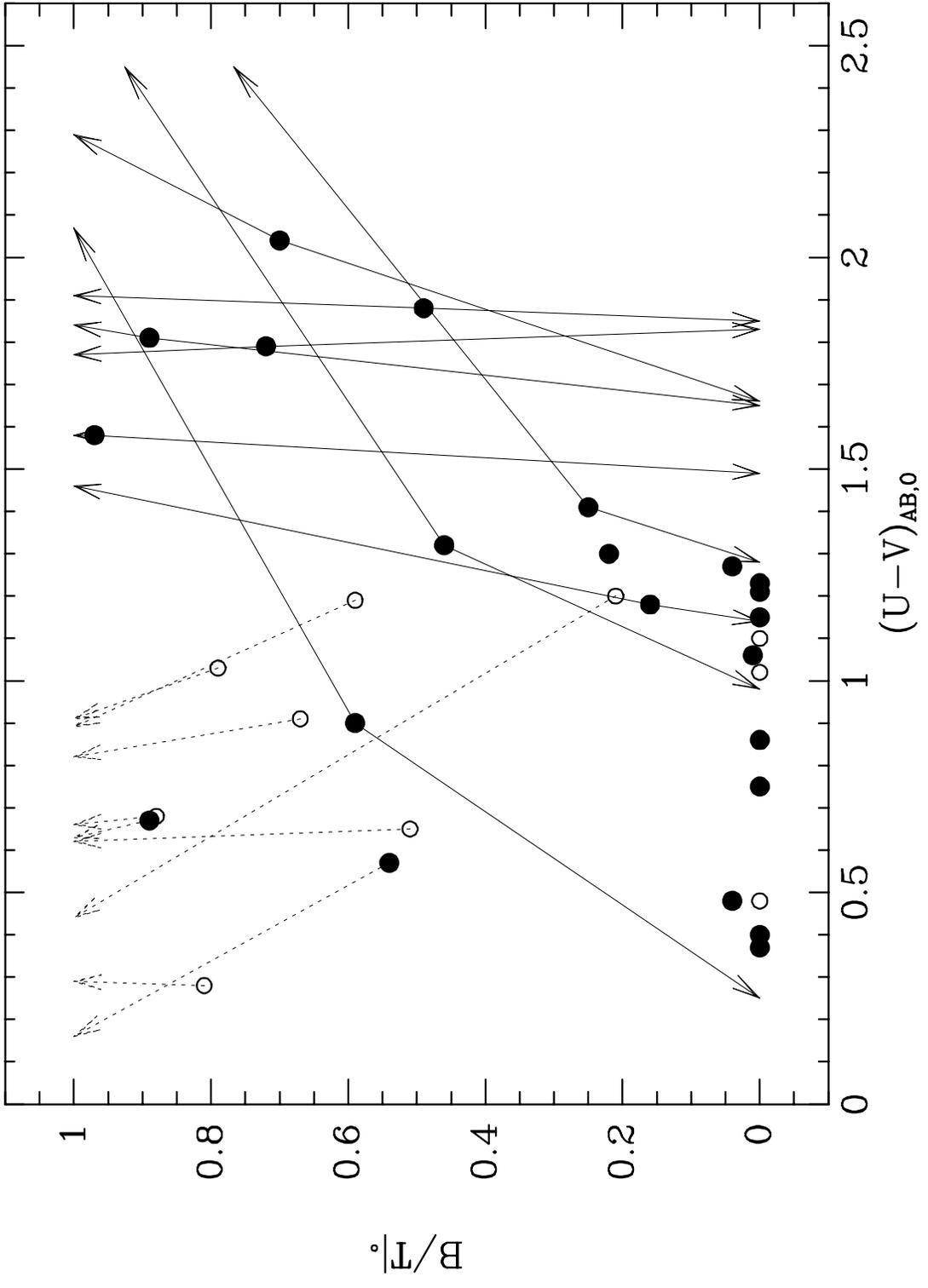

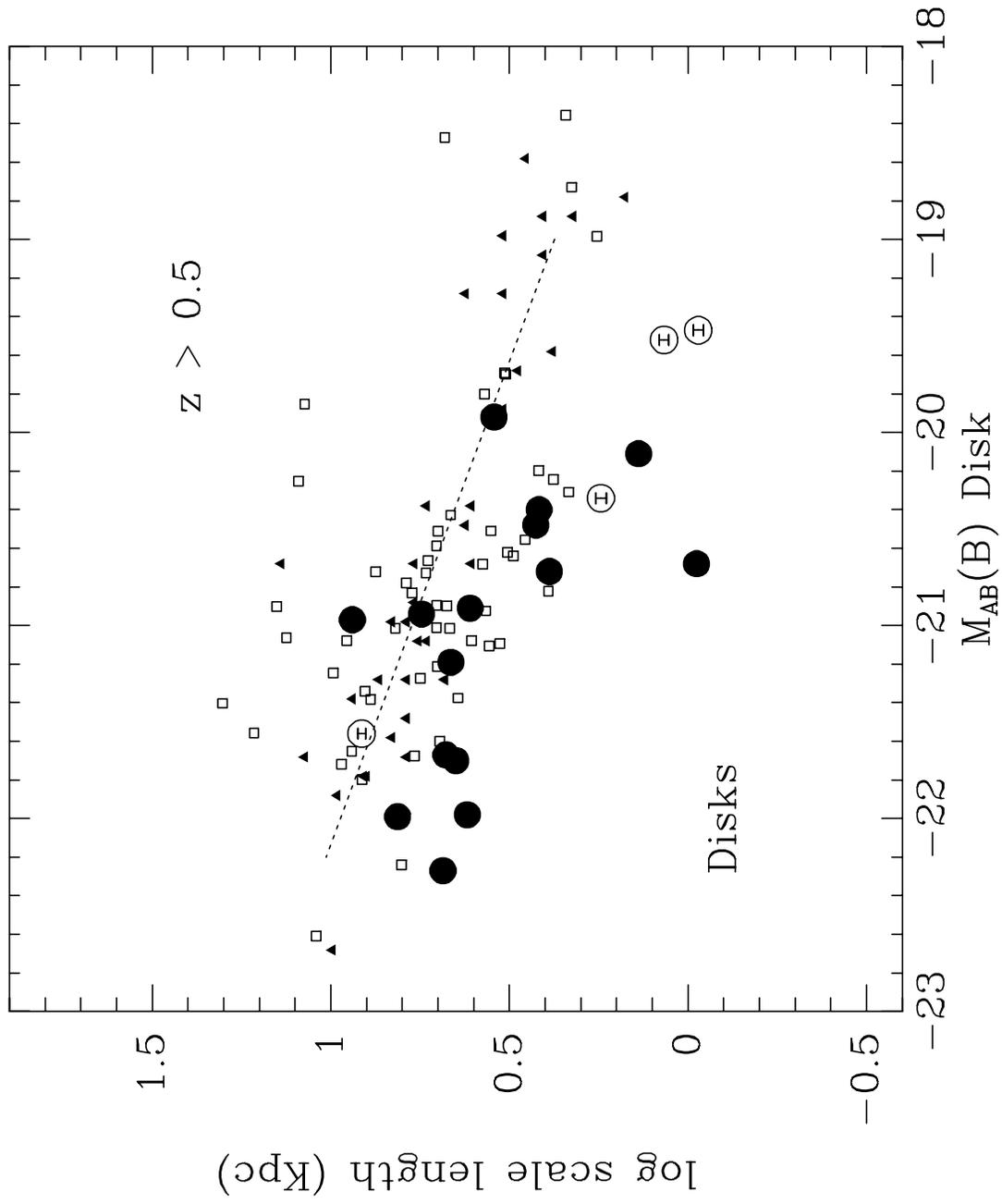

# CANADA-FRANCE REDSHIFT SURVEY: HST IMAGING OF HIGH-REDSHIFT FIELD GALAXIES

David Schade and S. J. Lilly
Department of Astronomy, University of Toronto, Toronto, Canada M5S 1A7

David Crampton
Dominion Astrophys. Obs., National Research Council of Canada, Victoria, V8W 4M6

F. Hammer, O. Le Fèvre & L. Tresse
DAEC, Observatoire de Meudon, 92195 Meudon, France

## ABSTRACT

*Hubble Space Telescope*[1] $B$ and $I$ images are presented of 32 galaxies with secure redshifts in the range $0.5 < z < 1.2$ from the Canada-France Redshift Survey, a complete sample of galaxies with $17.5 \leq I_{AB} \leq 22.5$. These galaxies exhibit the same range of morphological types as seen locally, i.e., ellipticals, spirals and irregulars. The galaxies look far less regular in the $B$ images (rest-frame ultraviolet) than at longer wavelengths, underlining the fact that optical images of galaxies at still higher redshift should be interpreted with caution.

Quantitative analyses of the galaxies yield disk sizes, bulge fractions, and colors for each component. At these redshifts, galaxy disks show clear evidence for surface brightness evolution. The mean rest-frame central surface brightness of the disks of normal late-type galaxies is $\mu_{AB}(B) = 20.2 \pm 0.25$ mag arcsec$^{-2}$, about 1.2 magnitudes brighter than the Freeman (1970) value. Some degree of peculiarity/asymmetry is measurable in 10 (30%) of the galaxies and 4 (13%) show clear signs of interaction/mergers. There are 9 galaxies (30%) dominated by blue compact components. These components, which appear to be related to star formation, occur most often in peculiar/asymmetric galaxies (some of which appear to be interacting), but a few are in otherwise normal galaxies. Thus, of the galaxies bluer than present-day $Sb$, one-third are "blue nucleated galaxies", and half are late-type galaxies with disks which are significantly brighter than normal galaxies at $z = 0$. Taken together, these two effects must be responsible for much of the observed evolution of the luminosity function of blue galaxies.

*Subject headings:* galaxies:evolution—galaxies:fundamental parameters

---

## 1. INTRODUCTION

The spatial resolution of *HST* images makes it possible to recognize the same features that are familiar from studies of nearby galaxies, in objects as distant as redshift $z = 1$, corresponding to a lookback time of $2/3\tau_o$ for $\Omega = 1$. *HST* imaging is thus a powerful tool that can help distinguish between competing explanations of field galaxy evolution, e.g., widespread merging (Broadhurst, Ellis and Glazebrook 1992), fading low-mass populations (Babul and Rees 1992), differential luminosity evolution, or any combination thereof (see Lilly, 1993 for a review).

Several studies of the field galaxy population with *HST* are in progress, but no clear picture has yet emerged (e.g., Mutz et al. 1994; Koo et al. 1994; Forbes et al. 1994; Griffiths et al. 1994a, b; Glazebrook et al. 1995). One of the difficulties in interpreting the data has been the lack of complete redshift data for the observed samples. Redshifts are essential since, for example, at $I_{AB} \sim 22.5$ galaxies span a factor of ten in redshift, from $z = 0.1$ to $z > 1$, and thus exhibit an enormous range of luminosity, scale size, and epoch. The morphologies of large numbers of faint galaxies revealed on the *HST* images can be investigated, but their interpretation relies on modelling and extrapolating data from the local population. The approach adopted in this *Letter* involves the analysis of a sample drawn from a complete redshift survey whose selection properties are well-determined. In this case, the luminosities, rest-frame colors, spectral energy distributions, and scale sizes in physical units are all known, allowing a more direct interpretation of the results.

The Canada-France Redshift Survey (CFRS) contains spectra of a statistically complete magnitude-limited sample of 943 objects with $17.5 \leq I_{AB} \leq 22.5$ (see Le Fèvre et al. 1995 CFRS II, Crampton et al. 1995 CFRS V, and references therein). Four *HST* pointings in F450W ($B$) and F814W ($I$) in the 0300+00 (Hammer et al. 1995 CFRS IV) and 1415+52 (Lilly et al. 1995b CFRS III) CFRS fields were obtained, with integration times of 6000s each, in 1994 November and December. The PC was pointed at random $z \sim 0.6$ galaxies, with the result that images were obtained for 45 galaxies with secure redshifts $0 < z < 1.2$, one QSO at $z = 1.6$, 8 stars and 3 objects for which the CFRS spectra did not yield secure spectroscopic identifications. The *HST* photometric zeropoints were adopted and found to be in good agreement (to $\pm 0.1$ mag) with the ground-based photometry (Lilly et al. 1995a CFRS I). Rest-frame $(U - V)_{AB}$ colors and $B$-band luminosities were derived from the observed magnitudes and redshifts, interpolating among the spectral energy distributions of Coleman, Wu, and Weedman (1980) as described by Lilly et al. 1995b (CFRS VI).

This *Letter* analyzes the images of the subsample of 32 galaxies with secure redshifts, $z > 0.47$. Of the three spectroscopically unidentified galaxies, two of them are likely to lie at $z > 0.5$ on the basis of tentative redshifts derived from their spectra and their colors (see Crampton et al. 1995 CFRS V) but these two galaxies have morphologies similar to those of the identified galaxies and their exclusion is unlikely to bias the sample significantly. The qualitative properties of the galaxies in our sample are discussed in §2. Quantitative analyses of the light profiles are described in §3 and a discussion of the results in §4. It is assumed that $H_o = 50$ km sec$^{-1}$ Mpc$^{-1}$ and

$q_\circ = 0.5$ throughout.

## 2. QUALITATIVE EVALUATION OF HIGH REDSHIFT GALAXIES

Figures 1-2 (plates XXX) show images of the 32 CFRS galaxies with $0.5 < z < 1.2$ observed with *HST*. At these redshifts, $I$ is equivalent to rest-frame $B$ or $V$ so that direct visual comparisons can be made with local galaxies. The $B$ band samples the rest-frame ultraviolet region (shortward of 2900 Å). The galaxies are arranged from the top left in order of rest-frame $(U - V)$ color (bluer toward the right) and rest-frame blue luminosity (decreasing downward). Identification and details of all the galaxies are listed in Table 1. Pseudo true-color images produced by combining the two images are shown in Figure 3 (plate XXX).

Many of the galaxies in Figures 1-3 have close counterparts in the Hubble Atlas of Galaxies (Sandage 1961) (e.g., c.f. 14.0393 to M101). Spiral structure is visually apparent in five galaxies and obvious irregular structure in several. Among those objects that appear to have normal disk-plus-bulge morphology, most have component colors consistent with what is found locally, i.e., redder bulge, bluer disk, (e.g., 14.1043, 14.0854, 14.0393). However, nine of the galaxies (e.g., 03.1540 and 14.0985) appear to have blue central concentrations. The frequency of asymmetric or irregular structure increases at low luminosity. Companions within 25 $h_{0.5}^{-1}$ kpc are seen in five cases. If merging galaxies are defined as those objects with high surface-brightness irregular structure, accompanied by low surface brightness asymmetric (tidal) features, then there are four cases of "mergers" (14.0972, 14.1139, 03.0488, 03.1540).

As anticipated, the galaxies look much less regular in the $B$ images (rest-frame ultraviolet). For example, 14.1043 appears to be a normal early-type spiral in the $I$ band image, whereas it shows two equal surface-brightness components in the $B$ frame; 03.0445 is unclassifiable in the $B$ image although it is a regular spiral in $I$. Many of the galaxies which look "irregular/peculiar" in the $B$ images would be characterized as normal in the $I$ images. This emphasizes the severe difficulties involved in interpreting optical images of galaxies at z> 1.

## 3. QUANTITATIVE ANALYSES

Two fundamental properties of a galaxy are its size and morphological type. These quantitites were estimated by fitting bulge-plus-disk models to "symmetric" $I$ images of the galaxies generated in the following way (see e.g., Elmegreen, Elmegreen, & Montenegro 1992). An "asymmetric image" was first constructed by rotating the original image about the galaxy center and subtracting this from the original image. The resulting frame was clipped to exclude positive structure less significant than $2\sigma$ and then subtracted from the original, giving an image from which strong asymmetric structures have been removed. A $\chi^2$ minimization procedure (Schade et al. 1995) was used to determine the best fit disk scale length, bulge effective radius, and surface brightness, inclination, and orientation for each component. Structural parameters were derived from the $I$ image and component colors were determined by scaling the surface brightness of the bulge and disk separately, holding structural parameters fixed, until a satisfactory fit to the $B$ image was achieved. The rest-frame colors and luminosities were determined for the bulge and disk components individually. Using these colors, the final bulge fraction ($B/T$) was defined in rest-frame $B$.

The best-fit model was subtracted from the original image and residual structure measured using indices which estimate the remaining flux. Defining $I_{ij}$ as the original image and $R_{ij}$ and $R_{ij}^{180}$ as the residual (original minus model) image and its counterpart after a 180 degree rotation:

$$R_T = \frac{\sum \frac{1}{2}|R_{ij} + R_{ij}^{180}|}{\sum I_{ij}} \qquad R_A = \frac{\sum \frac{1}{2}|R_{ij} - R_{ij}^{180}|}{\sum I_{ij}}.$$

The total and asymmetric indices $R_T$ and $R_A$ (computed within a radius of 5 kpc) are approximately equal to the total and asymmetric residual flux respectively, expressed as a fraction of the total galaxy flux. An object was classified "asymmetric" if $R_T + R_A > 0.14$. In practical terms, most objects with $R_A > 5\%$ of the galaxy flux are classified as asymmetric.

## 4. RESULTS

### 4.1. Color versus morphology

The local galaxy population exhibits well-defined relations between integrated color and morphological type (e.g., Buta et al. 1994), due largely to the varying proportions of bulge (red) and disk (bluer) luminosity as a function of Hubble type. Since the bulge/total luminosity ($B/T$) correlates with Hubble type, a variation of $B/T$ with color is expected. Figure 4 shows the color-morphology relation for the 32 high redshift galaxies. The symbols indicate integrated ($U - V$) restframe colors and $B/T$, and the arrows show the colors of the corresponding bulge and disk components. Open symbols indicate those galaxies classified as asymmetric/peculiar. The expected galaxy sequences are present in this diagram. Early-type galaxies populate the quadrant $(U - V)_{AB,o} > 1.4$, $B/T > 0.5$. There is a group of disk-dominated galaxies with $(U - V)_{AB,o} < 1.4$, $B/T < 0.5$. Note, however, that there are nine galaxies with decomposed colors (dashed arrows) indicating blue, compact components unlike those in normal galaxy sequences. Although we refer to decomposition into "bulges", these components are sometimes not centered in the galaxies and so we refer to these galaxies as "blue nucleated galaxies" (BNGs). The association of asymmetric/peculiar structure (open symbols) with the BNG phemomenon (dashed arrows) is clear.

### 4.2. Blue nucleated galaxies

Among the nine BNGs, three show clear indications of interaction/merging based on the presence of high surface brightness, irregular structure in addition to the presence of low surface brightness asymmetric (presumably tidally-induced) structure. Seven of the nine BNG objects are asymmetric/peculiar as measured by our objective residual indices. A test of these correlations using a four-fold table (Sachs 1984) indicates that the associations of merging/interaction and asymmetric structure with the BNG phenomenon are significant at the 97% and 99.9% level of confidence respectively. These associations suggest that these are not simply blue bulges; it seems more likely that they are associated with starburst phenomena. Measurements of the O[II]/H$\beta$ and O[III]/H$\beta$ ratios from our spectra indicate that BNGs are the result of star formation rather than an AGN-related phenomenon.

### 4.3. Disk surface brightness

Figure 5 shows the relationship between disk luminosity and scale length. The solid circles show disks of galaxies with $B/T < 0.5$ and open circles those of galaxies with $B/T < 0.75$. Asymmetric galaxies, as defined in §3, have been excluded. For comparison, equivalent data from Kent (1985, small open squares) and van der Kruit (1987, small open triangles) have also been plotted (converted to $B_{AB} = B - 0.2$). The dotted line represents the Freeman (1970) constant surface brightness law. If the $z = 0$ locus defined by Freeman's law is correct (and it is supported by the two datasets shown), then this sample of disk galaxies at $z > 0.5$ shows strong evolution in rest-frame central surface brightness. The mean rest-frame value, excluding irregular and bulge-dominated galaxies and corrected for inclination, is $\mu_{AB}(B) = 20.2 \pm 0.25$ mag arcsec$^{-2}$, approximately 1.2 magnitudes (nearly 5$\sigma$) brighter than the Freeman value. Galaxies showing obvious spiral structure have a mean $\mu_{AB}(B) = 20.1 \pm 0.2$, small featureless objects have $\mu_{AB}(B) = 19.73 \pm 0.5$, and irregular objects (where the meaning of the fitted disk central surface brightness may be questionable) have $\mu_{AB}(B) = 20.55 \pm 0.8$. Thus, there is no significant difference in $\mu_{AB}(B)$ among these subsamples and therefore no evidence that the observed surface brightness evolution is dependent upon a galaxy's membership in an easily recognized Hubble class, nor even upon its degree of regularity. It should be noted that surface brightness measurements are independent of $q_o$.

### 5. DISCUSSION

Three conclusions follow from the present observations. First, the galaxy population at $z \sim 0.75$ is broadly similar to the local galaxy population. The same range in morphological properties is apparent, and all present-day galaxy types are represented. The basic similarity of the high-redshift and local populations indicates that exotic evolutionary processes do not dominate luminous galaxy evolution.

Second, in addition to populations that are normal in the context of the color-morphology relation, there exist blue nucleated galaxies, BNGs, (30% of the sample) which are predominantly associated with asymmetric structure and in a few cases with obvious interactions/mergers. Because the BNGs represent 35% of the blue galaxy population whose luminosity function is known to be strongly evolving (Lilly et al. 1995b) they must play a significant role in that evolution.

Finally, the average central surface brightness of the disks of normal late-type galaxies ($B/T < 0.5$), is higher by $1.2 \pm 0.25$ magnitude/arcsec$^{-2}$ than in the local population. If it is assumed that disk sizes have been constant, then this represents the second major component of the observed evolution in the luminosity function of blue galaxies (since they represent 50% of the blue galaxies). The surface brightness enhancement is seen in galaxies dominated by regular bulge-plus-disk structure with clear spiral arms, and also in irregular and compact objects (15% of the blue galaxies are asymmetric/peculiar and are not BNG's but show high disk central surface brightness). These results imply that the majority of disk galaxies at $z \sim 0.75$ are forming stars over much of the extent of their disks at a rate of 2-3 times higher than the local average rate.

An order of magnitude increase in sample size is required to provide definitive estimates of the distribution of morphological types and the frequency of bars, spiral structure, merging, and the BNG phenomomen. Even the small sample analysed here, however, illustrates the enormous power of *HST* resolution, combined with necessary redshift information, to allow us to see the distant universe in terms that are comparable to the way we observe the universe locally.

We thank Ted von Hippel for sharing data which helped us characterise the WFPC2 point-spread function. We also acknowledge the indirect contribution to this work of all those associated with the *HST* project. This work was supported financially by NSERC of Canada.

---

Fig. 1.— WFPC2 images of 32 galaxies at $0.5 < z < 1.2$ in the F814W filter ($\sim I$). The galaxies are arranged such that objects with increasingly blue (rest-frame) colors are on the right, higher luminosities at the top. The galaxies can be identified by their position in this mosaic given in Table 1.

Fig. 2.— The set of 32 galaxies at $0.5 < z < 1.2$ in the F450W ($\sim B$) filter. The order is the same as in Fig. 1. Many galaxies that are apparently normal in the $I$-band image would clearly *not* be considered normal based on their appearance in the $B$ (restframe ultraviolet) image.

Fig. 3.— Pseudo true-color images of CFRS galaxies constructed by combining the $I$ and $B$ WFPC2 images. The order is the same as in Fig. 1.

Fig. 4.— The color-morphology relation for high-redshift galaxies. The symbols indicate the integrated rest-frame color and $B/T$ for each galaxy. The arrows indicate the corresponding component color (the bulge color of a galaxy is the colour where its arrowed line intersects the line $B/T = 1$, the disk color where its arrowed line intersects $B/T = 0$). Open circles are objects with asymmetric structure (see text). Blue nucleated galaxies are shown with dashed arrows.

Fig. 5.— The relation between disk scale length and disk luminosity for the normal galaxies in the sample. The dashed line indicates the Freeman (1970) constant surface-brightness relation ($\mu(B) = 21.65$ mag arcsecond$^{-2}$). Samples from Kent (1985) and van der Kruit (1987) are plotted as small squares and triangles respectively. Large solid symbols indicate objects with $B/T < 0.5$ and large open symbols, $0.5 < B/T < 0.75$

TABLE 1
CFRS Galaxies at $z > 0.5$

| CFRS No. | z | $M_B(AB)$ | $(U-V)_{AB}$ | B/T | Log $h$ | $R_T$ | $R_A$ | X Y | Remarks |
|---|---|---|---|---|---|---|---|---|---|
| 03.0445 | 0.530 | −21.48 | 1.23 | 0.00 | 0.68 | 0.06 | 0.05 | 3 7 | spiral |
| 03.0480 | 0.608 | −19.72 | 0.37 | 0.00 | 0.54 | −0.01 | −0.19 | 6 1 | |
| 03.0485 | 0.606 | −19.91 | 0.37 | 0.00 | 0.14 | 0.08 | −0.04 | 6 2 | |
| 03.0488 | 0.607 | −20.07 | 0.48 | 0.00 | 0.23 | 0.16 | 0.14 | 5 1 | M/I,A |
| 03.0523 | 0.651 | −20.91 | 0.68 | 0.88 | 0.58 | 0.18 | 0.19 | 4 5 | BNG,A |
| 03.0528 | 0.714 | −21.39 | 1.32 | 0.46 | 0.61 | 0.06 | 0.02 | 3 5 | |
| 03.0560 | 0.697 | −21.32 | 1.81 | 0.89 | ... | 0.04 | −0.01 | 2 5 | |
| 03.0579 | 0.660 | −20.28 | 1.15 | 0.00 | 0.43 | 0.01 | −0.00 | 3 1 | |
| 03.0595 | 0.606 | −20.48 | 1.21 | 0.00 | −0.02 | 0.08 | −0.03 | 3 2 | |
| 03.0599 | 0.481 | −21.01 | 1.30 | 0.22 | 0.75 | −0.03 | −0.08 | 3 4 | |
| 03.0717 | 0.607 | −21.54 | 1.27 | 0.04 | 0.65 | 0.03 | 0.01 | 3 6 | spiral |
| 03.1531 | 0.715 | −20.96 | 1.18 | 0.16 | 0.94 | 0.00 | −0.03 | 3 3 | |
| 03.1540 | 0.690 | −21.34 | 1.03 | 0.79 | 0.86 | 0.10 | 0.05 | 4 4 | spiral,bar,BNG,M/I,A |
| 14.0147 | 1.181 | −21.80 | 1.06 | 0.01 | 0.81 | 0.05 | −0.05 | 4 7 | |
| 14.0207 | 0.546 | −22.67 | 2.04 | 0.70 | ... | 0.02 | 0.01 | 1 8 | |
| 14.0393 | 0.602 | −21.83 | 0.48 | 0.04 | 0.62 | 0.06 | 0.02 | 5 7 | spiral |
| 14.0485 | 0.655 | −20.20 | 0.75 | 0.00 | 0.42 | 0.03 | 0.02 | 5 2 | |
| 14.0846 | 0.989 | −21.39 | 1.20 | 0.00 | 0.66 | 0.13 | 0.03 | 5 4 | spiral,A |
| 14.0854 | 0.992 | −21.98 | 1.58 | 0.97 | ... | 0.08 | 0.00 | 2 7 | |
| 14.0962 | 0.753 | −21.73 | 1.88 | 0.49 | 0.66 | 0.04 | 0.03 | 2 6 | |
| 14.0972 | 0.674 | −21.03 | 0.65 | 0.51 | 0.46 | 0.18 | 0.15 | 5 5 | BNG, M/I,A |
| 14.0985 | 0.807 | −20.98 | 0.57 | 0.54 | ... | 0.05 | 0.05 | 6 4 | BNG |
| 14.1012 | 0.479 | −20.54 | 0.91 | 0.67 | ... | 0.08 | 0.06 | 4 2 | BNG,A |
| 14.1028 | 0.988 | −21.60 | 1.79 | 0.72 | 0.27 | 0.07 | 0.06 | 1 7 | |
| 14.1037 | 0.549 | −20.73 | 1.02 | 0.00 | 0.36 | 0.11 | 0.07 | 4 3 | A |
| 14.1042 | 0.722 | −21.44 | 1.19 | 0.59 | 0.01 | 0.14 | 0.12 | 4 6 | BNG,A |
| 14.1043 | 0.641 | −22.39 | 1.41 | 0.25 | 0.69 | 0.04 | 0.02 | 2 8 | |
| 14.1087 | 0.660 | −21.16 | 0.67 | 0.89 | ... | 0.03 | 0.01 | 5 6 | BNG |
| 14.1136 | 0.640 | −20.46 | 0.28 | 0.81 | −0.01 | 0.13 | 0.10 | 6 3 | BNG,A |
| 14.1139 | 0.660 | −22.45 | 1.20 | 0.21 | 1.03 | 0.18 | 0.13 | 4 8 | BNG, M/I,A |
| 14.1189 | 0.753 | −20.52 | 0.86 | 0.00 | 0.39 | 0.03 | 0.03 | 5 3 | |
| 14.1258 | 0.645 | −20.23 | 0.90 | 0.59 | ... | 0.01 | −0.01 | 4 1 | |

Note.— Absolute magnitudes assume $H_0 = 50$ km sec$^{-1}$ Mpc$^{-1}$, $q_0 = 0.5$. Scale length $h$ is in kpc. In Remarks "A" indicates asymmetric structure, "M/I"= merger/interaction, "BNG"= blue nucleated galaxy. "Spiral" indicates obvious spiral structure. X and Y give the position of the object from the origin at bottom left in figures 1-3.